\documentclass{article}
\usepackage[T1]{fontenc} 
\usepackage[utf8]{inputenc} 
\usepackage{ismir,amsmath,cite,url}
\usepackage{graphicx}
\usepackage{pgfplots}
\usepackage{color}
\usepackage{multirow}

\usepackage{lineno}
\usepackage[bookmarks=false,colorlinks,linkcolor=blue]{hyperref} 

\title{Can LLMs "Reason" in Music? An Evaluation of LLMs' Capability of Music Understanding and Generation}

\multauthor
{Ziya Zhou$^{1,2}$ \hspace{1cm} Yuhang Wu$^2$ \hspace{1cm} Zhiyue Wu$^3$} 
{\bfseries{Xinyue Zhang$^2$ \hspace{0.5cm} Ruibin Yuan$^{1,2}$ \hspace{0.5cm} Yinghao Ma$^{2,4}$} \\ {\textbf{Lu Wang}$^3$ \hspace{0.5cm} \textbf{Emmanouil Benetos}$^4$ \hspace{0.5cm} \textbf{Wei Xue}$^1$ \hspace{0.5cm} \textbf{Yike Guo}$^1$}\\
$^1$ AIS, The Hong Kong University of Science and Technology\\
$^2$ Multimodal Art Projection\\
$^3$ Shenzhen University\\
$^4$ C4DM, Queen Mary University of London\\
{\tt\small zzhoucp@connect.ust.hk, weixue@ust.hk, yikeguo@ust.hk}
}

\def\authorname{Z. Zhou, Y. Wu, and Z. Wu}

\begin{document}

\maketitle
\begin{abstract}
Symbolic Music, akin to language, can be encoded in discrete symbols. Recent research has extended the application of large language models (LLMs) such as GPT-4 and Llama2 to the symbolic music domain including understanding and generation. Yet scant research explores the details of how these LLMs perform on advanced music understanding and conditioned generation, especially from the multi-step reasoning perspective, which is a critical aspect in the conditioned, editable, and interactive human-computer co-creation process. 
This study conducts a thorough investigation of LLMs' capability and limitations in symbolic music processing. We identify that current LLMs exhibit poor performance in song-level multi-step music reasoning, and typically fail to leverage learned music knowledge when addressing complex musical tasks. An analysis of LLMs' responses highlights distinctly their pros and cons.
Our findings suggest achieving advanced musical capability is not intrinsically obtained by LLMs, and future research should focus more on bridging the gap between music knowledge and reasoning, to improve the co-creation experience for musicians.
\end{abstract}

\section{Introduction}\label{sec:introduction}

%


Large language models (LLMs), such as GPT-4, harness the power of deep learning to produce human-like text. These models, trained on vast datasets of textual content, have notably propelled advancements in natural language processing (NLP). They excel in complex language understanding and generation tasks including translation, sentiment analysis, question answering, and summarization, showcasing their reasoning capability with sophistication.

Large language models (LLMs), initially pre-trained on extensive textual corpora, can assimilate general linguistic patterns and structures. They are subsequently fine-tuned with domain-specific data, such as code and mathematical symbols, to enhance the adaptation to specific tasks. This refinement allows LLMs' proficiency to more accurately manage domain-specific terminology and complicated challenges like multi-step reasoning. Music Reasoning refers to the ability to estimate the varying harmonies, keys, rhythms, and other musical elements that are not explicitly annotated in a piece of music and are significant for music themes, progression, and styles~\cite{yuan_chatmusician_2024}. The analogy between the reasoning process in music and mathematics suggests their structural similarities. Both disciplines fundamentally rely on patterns: music in rhythms, scales, and chord progressions, while mathematics involves sequences, symmetries, and geometric configuration. Moreover, music theory utilizes mathematical concepts to articulate intervals between pitches, chord structures, and the rhythmic temporal division ~\cite{garland1995math, wright2009mathematics}, underscoring the intrinsic reasoning nature of the musical components. 

Music can be represented as sequences of symbols such as MIDI or ABC notation, rendering it suitable for processing by LLMs, which excel in long-context understanding and multi-step reasoning. These models are capable to dissect and generate intricate musical patterns encompassing melodic, harmonic, and rhythmic structures. LLMs also play a pivotal role in enhancing interactive music generation systems, where user inputs tailor the model's output, enriching the composing experience. While previous studies~\cite{yuan_chatmusician_2024, ding2024songcomposer, liang_bytecomposer_2024} have investigated LLMs in music tasks, detailed interpretations of the process remains less explored. This paper conduct an evaluation of four LLMs, GPT-4~\cite{achiam2023gpt}, Gemma-7B-it~\cite{team2024gemma}, Llama2-7B-chat~\cite{touvron2023llama}, and Qwen-7B-chat~\cite{bai2023qwen}, assessing their capabilities on tasks related to symbolic music understanding and generation:

\begin{itemize}
    \item Music Understanding: 1) Music theory exercise; 2) Motif extraction; 3) Musical form extraction.
    \item Music Generation: 1) Chord-conditioned music generation; 2) Melody harmonization; 3) Musical-form-and-motif-conditioned music generation
\end{itemize}

\begin{figure*}
\centering
\includegraphics[width=2.1\columnwidth]{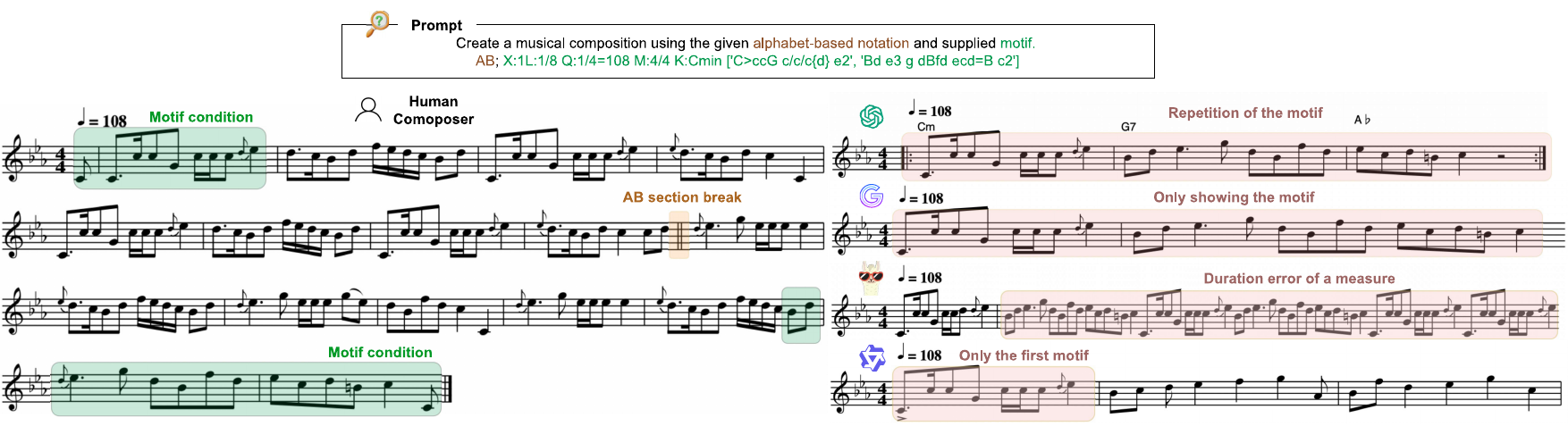}
\caption{A comparison of different LLMs' responses with the same instruction of the musical-form-and-motif-conditioned task as the input. The ABC notation contained in the response is extracted and displayed as scores the quality of all responses is marked with diverse symbols.}
\label{fig:different_models_responses}
\end{figure*}

The task of "Musical Form \& Motif Conditioned Music Generation" as described in Chatmusician~\cite{yuan_chatmusician_2024} involves generating music that adheres to detailed prescribed conditions like form and motif. Figure \ref{fig:different_models_responses} illustrates this process: The prompt's green text specifies conditional constraints including the musical form, motif, and some musical elements (key, time signature, etc.). Under the prompt, the left sheet presents the human composer's work. The right sheets show ABC notations from different models alongside the reference. The Gemma-7B-it model merely replicates the provided motif, adding no new elements. Similarly, GPT-4 simply repeats the given condition. Qwen-7B-chat and Llama-7B-chat include correct musical elements and the motif but fail to capture the musical form \textit{"AB"} and maintain the duration of a measure.


The main contributions of our paper are as follows: (1) we provide multi-step prompt engineering and explore how LLMs exhibit their reasoning capabilities with multi-step instructions in music understanding and generation tasks. (2) we assess four major LLMs on various symbolic music tasks, analyzing their reasoning in ABC sequences through quantitative statistical results and qualitative human assessment, including error analysis. The examples, hand-crafted prompts, and codes of data preprocessing are available at \href{https://github.com/SylviaZiyaZhou/LLMs\_music\_reasoning}{github}.

\section{Related Work}
In this section, we summarize related works from two perspectives. First, we introduce previous studies on LLMs in the symbolic music domain, explaining their performance and evaluation methods in music understanding and generation tasks. Then, we discuss the application of LLMs in reasoning math problems and controllable creative text generation, highlighting similarities between the reasoning processes in music and math and the conditioned, open-ended nature of both music and text generation.

\subsection{LLMs in Symbolic Music Domain}
This subsection reviews the application of LLMs in the symbolic music domain. Previous studies have focused on adapting LLMs for music understanding and generation. 
Chatmusician~\cite{yuan_chatmusician_2024} uses continual pre-training and fine-tuning on LLaMA2 to understand and generate ABC notation music, without specialized music structures or tokenizers. SongComposer~\cite{ding2024songcomposer} collects a song pretraining dataset including lyrics, melodies and paired lyrics-melodies, employing 10K crafted QA pairs to enable LLMs to perform multiple music-related tasks such as lyric-to-melody conversion and song continuation. MusicAgent~\cite{yu2023musicagent} integrates various music tools into a single system, though it lacks interaction among these tools. 
Most approaches view music creation as a linear process, which diverges from the multi-step approach humans use, limiting their applicability for generating creative works.  To mimic human creative processes,  ByteComposer~\cite{liang_bytecomposer_2024} employs a four-step method to replicate the creative workflow of human composers: conception analysis, draft Composition, self-evaluation and modification, and human aesthetic selection. 
And designs an interactive agent system consisting of expert, generator, voter, and memory modules. What's more, they construct supervised fine-tuning data covering tasks of basic music theory conception, control code generation, music score evaluation and next-step planning. Despite being a significant step towards multi-step music creation with LLMs, it lacks a detailed discussion on the limits of LLMs at each stage.

\subsection{Reasoning and Controllable Generation with LLMs}
"Reasoning" in NLP involves integrating various knowledge sources or contexts to generate new assertions, events, or actions~\cite{yu2023reasoningsurvey}.  This process often breaks complex questions into sequential steps~\cite{hao2023reasoningLM}.  Techniques such as Chain-of-Thoughts (CoT)~\cite{wei2022chain, kojima2022zeroshotlearners} have shown effectiveness in addressing complex reasoning tasks, particularly in mathematics. The Program-of-Thoughts approach improves upon CoT by using language models to generate text and code, enhancing math problem-solving performance~\cite{chen2022program}. Plan-and-Solve (PS) Prompting, a zero-shot technique, outperforms zero-shot CoT significantly, exceeds Zero-shot Program-of-Thoughts, and matches 8-shot CoT in math reasoning~\cite{wang2023plan}.

While music and mathematics share similarities, it is crucial to recognize that music is not as deterministic. In controllable music generation, despite given chords, motifs, and forms, unpredictable elements still significantly affect the quality of the music, similar to controllable text generation. Zhang et al.\cite{zhang2023textgensurvey}, identify three types of control conditions: semantic, structural, and lexical. Semantic controls refer to content control such as sentiment~\cite{chen2019sentiment, dathathri2019plug} or topic~\cite{khalifa2020distributional, tang2019topic}, resembling style and emotion in music. Structural control involves shaping the structure of the generated text, such as setting a story's framework or using data from tables or graphs as input, similar to specifying musical forms for generation~\cite{puduppully2019contentplan, ribeiro2020investigating}. Lexical controls manage vocabulary usage, ensuring specific keywords appear, akin to using musical chords and motifs as guidelines. LLMs are extensively applied in diverse controllable and creative generation tasks ~\cite{simon2022tattletale, xie2024creating, zhang2024llm}. These systems' abilities in long-context and multi-step generation under predefined conditions are examined, though such analyses are rarely applied in the music domain.  



\section{Methodology}\label{sec:typeset_text}
\subsection{Datasets}
In this paper, we incorporate six tasks covering from music understanding to generation. The data is collected from \textit{MusicPile} and \textit{MusicBench} in ChatMusician~\cite{yuan_chatmusician_2024}. The statistics of the dataset we use are shown in Table \ref{tab:statistics}. Each model can support the maximum length of tokens of each task. 
 
\begin{table*}[!ht]
 \centering
 \begin{tabular*}{0.765\linewidth}{|l|l|l|}
  \hline
  Tasks & Numbers & Max/Avg. tokens \\
  \hline
  Music theory exercise & 367 & 733/103.56 \\
  Motif extraction (ME) & 2470 & 1165/194.28 \\
  Musical form extraction (MFE) & 483 & 650/187.35 \\
  Chord-conditioned generation (CCG) & 1721 & 283/94.83 \\
  Melody harmonization (MH) & 355 & 551/166.03 \\
  Musical-form-and-motif-conditioned generation (MFMC) & 4881 & 285/53.82 \\
  \hline
 \end{tabular*}
 \caption{Statistics of each task. The number of items and the max and average length of tokens are provided.}
 \label{tab:statistics}
\end{table*}

\subsection{Prompt Engineering}
Before examining each LLM's task performance, we conducted preliminary tests to verify their understanding of the relevant musical concepts. These tests confirmed that all models possess foundational knowledge of the six music tasks assessed in this study.

\begin{figure}
\centering
\includegraphics[width=1\columnwidth]{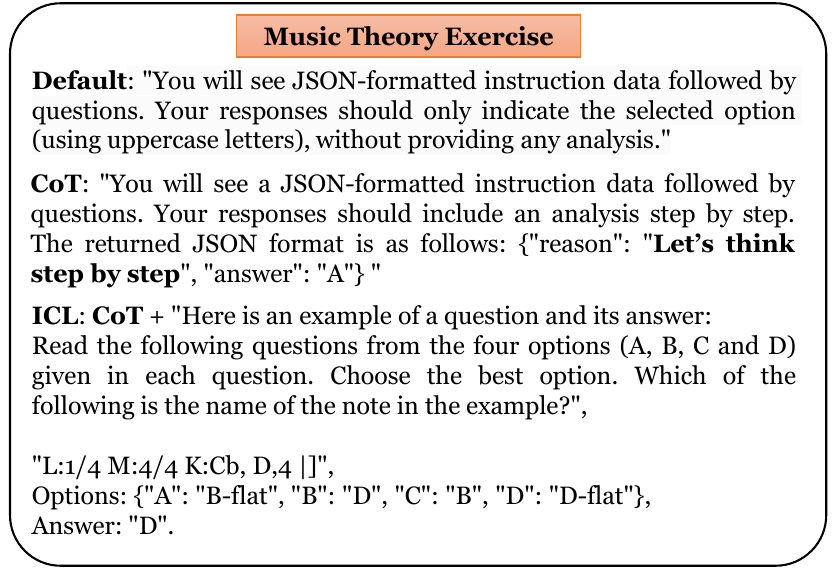}
\caption{A prompt example of the music theory exercise in different modes.}
\label{fig:Prompt_examples_theory}
\end{figure}

\begin{figure}
\centering
\includegraphics[width=1\columnwidth]{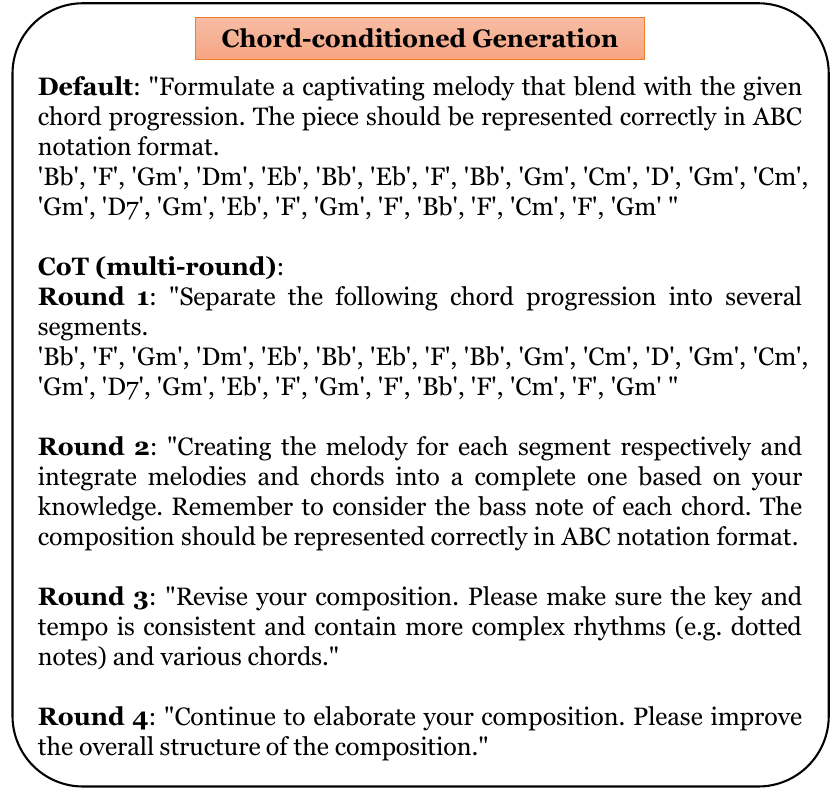}
\caption{A prompt example of the chord-conditioned generation in different modes. }
\label{fig:Prompt_examples_generation}
\end{figure}

We employed two prompt modes in our experiments of all tasks, \textbf{Default} and \textbf{Chain-of-Thoughts (CoT)}. Default mode means forcing the model to respond without any analysis. Additionally, for music theory exercises, to make the model better understand the questions and options, and return the answer in a unified format, we also include the \textbf{In-Context-Learning (ICL)} mode by adding some question-answer pairs as examples shown to the models in the prompt. Taking the task of music theory exercises understanding as an example, three modes of prompts as the prefix of inputs followed by each item in the datasets are shown in Figure~\ref{fig:Prompt_examples_theory}. Different from the music theory exercise, we specifically design prompts to support a multi-round chat conversation with LLMs for the generation tasks. Figure~\ref{fig:Prompt_examples_generation} shows an example of a four-round prompt set of chord-conditioned generation. We invite graduates who majored in music composition to write down their multi-step thoughts when completing the generation tasks involved in this paper. We summarize the common steps of all answers, adapt them to the prompt set, and make sure LLMs can understand or at least intend to follow the instructions. An example of GPT-4's response to the instruction in Figure~\ref{fig:Prompt_examples_generation} is shown on the website\footnote{https://github.com/SylviaZiyaZhou/LLMs\_music\_reasoning/blob/main/\\CoT\_music\_generation\_GPT4\_response.pdf}.

\subsection{Pre-processing Responses}\label{sec:pre-processing}
The responses of models are supposed to have correct ABC notations, but it may have certain syntax or formatting issues, and some outputs may even contain a large amount of natural language.
We select the main features of ABC notation including field names and bar line symbols to help us extract the ABC sequence. 
If the extracted ABC sequence can be rendered into MIDI files using Music21\footnote{https://web.mit.edu/music21} successfully and can be later rendering into audio file using midi2audio\footnote{https://pypi.org/project/midi2audio}, we consider it capable of producing valid ABC notation.

\subsection{Multi-step Reasoning Analysis}
In order to compare each model's reasoning capability on both understanding and generation tasks, we first conduct a subjective assessment to evaluate how different models' reasoning processes influence their performances. Participants are all familiar with basic music theory and can understand each task as well as the ABC notation. Secondly, based on the results of the subjective assessment, we further perform an error analysis in detail to show the intermediate answers during the reasoning process of each model.     

\subsubsection{Human Assessment Pipeline}
In this section, we will provide a detailed description of our subjective experiments on four popular and open-source LLMs, including Gemma-7B-it, Llama2-7B-Chat, GPT-4, and Qwen-7B-chat. 
We ask the participants to evaluate to what extent the model understands the instructions and correctly answer the questions in the understanding tasks, and to what extent the responses contain the conditions and make creative works in the generation tasks. Specifically, the questions in the human assessment are as follows:


\begin{itemize}
\item For both understanding and generation tasks: 1) To what extent does the model understand and follow the instructions? 
\item Specifically for the understanding tasks: 1) To what extent does the model correctly answer the question? 2) To what extent does the model reason like human beings?
\item Specifically for the generation tasks: 1) AB test: please choose the better one
between a pair of music excerpts by considering their "Musicality"; 2) To what extent does the model contain the conditions?
\end{itemize}

Except for the AB test in the generation task, each question should be rated in a scoring range from 0 to 10 points. We invited music experts who are familiar with ABC notations as the participants in the human assessment, ensuring that each item was evaluated by at least two experts. 

\section{Evaluation Results}
In this section, we provide the evaluation results based on the methodology we discussed in the last section. The quantitative results include the correctly parsing rate of ABC notation in the generation tasks, and the accuracy of music theory exercises. The qualitative results include the statistical analysis of human assessment and the detailed error analysis. Due to space limitation, we provide the examples at \href{https://github.com/SylviaZiyaZhou/LLMs\_music\_reasoning}{github} and the online links of the corresponding files will be attached in the illustration.

\subsection{Quantitative Results}
\begin{figure}[thbp!]
    \centering
    \begin{minipage}[t]{1.0\linewidth}
    \centering
        \begin{tabular}{@{\extracolsep{\fill}}c@{}@{\extracolsep{\fill}}}
            \includegraphics[width=\linewidth]{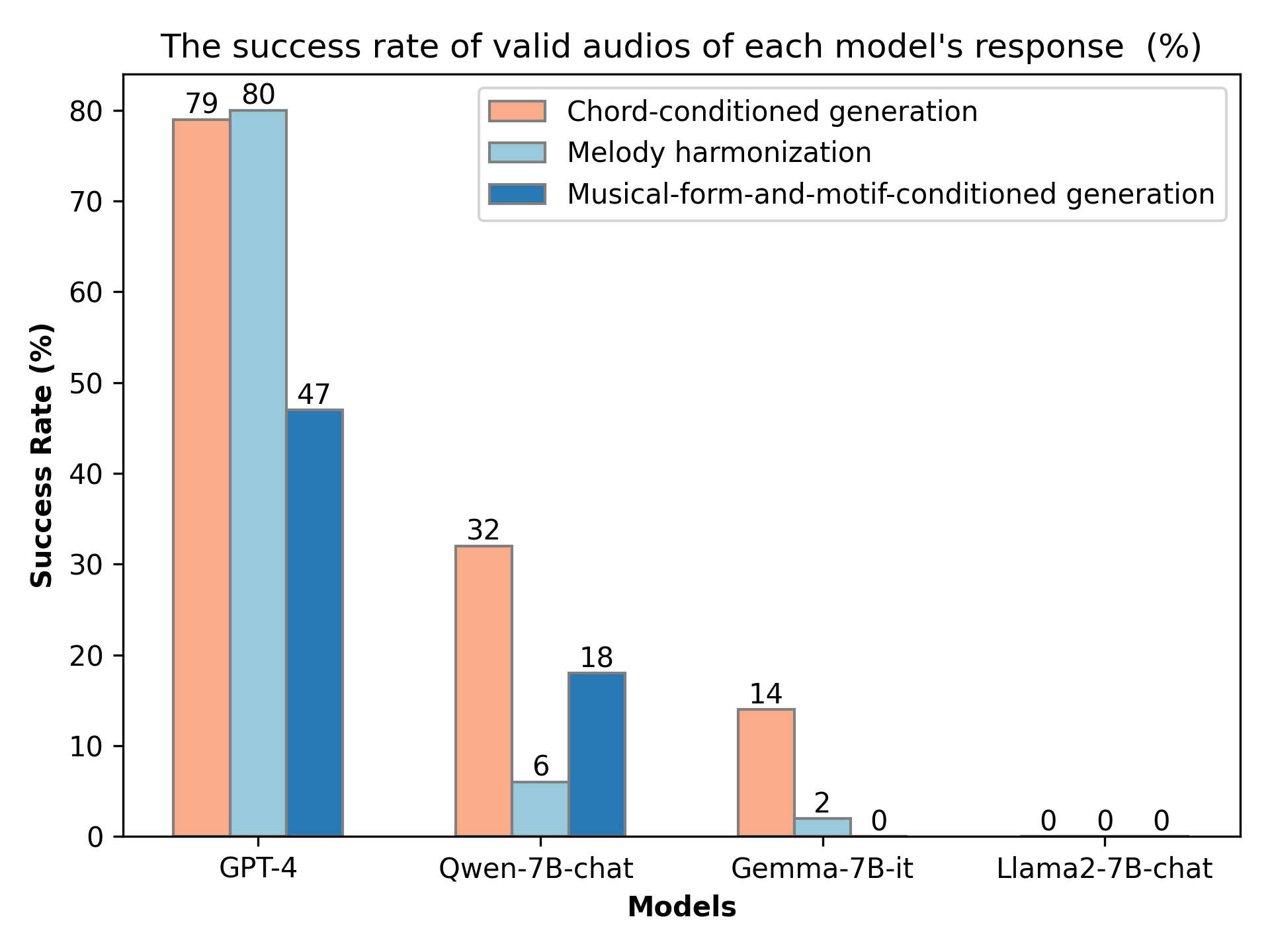}\\
           
        \end{tabular}
    \end{minipage}
   \caption{The success rate of rendering audio from each LLM's responses in the music generation tasks.}
    \label{fig:responses_ABC_statistics}
 \end{figure}

Figure~\ref{fig:responses_ABC_statistics} shows the success rate of rendering valid audio from each LLM's responses under different generation tasks. The pre-processing methodology is introduced in Section~\ref{sec:pre-processing}. Except for GPT-4, the other three models all have an audio generation rate of less than 50\%, finding it difficult to generate the correct ABC notation format to be converted into audio. 

Table~\ref{tab:music_bench_results} displays the accuracy of the music theory exercises in three modes. The reason why some models have an accuracy rate below 25\% in multiple-choice questions with four options is that most of their responses seek additional information about the questions rather than answering them. Gemma-7B-it has a comparable performance with GPT-4 in the \textit{Reason.} subset in the \textit{Default} mode even with a much smaller model size. However, CoT and ICL modes, which significantly improve the GPT-4's performance, show very limited effect or even deficiency in other models. This may inspire us to reconsider the utilization of classical CoT and ICL approaches in solving music tasks.

\begin{table}[!ht]
 \centering
 \begin{tabular}{|c|c|c|}
  \hline
  Model (and Mode) & \textit{Know.} (\%) & \textit{Reas.} (\%) \\
  \hline
  GPT-4 (Default) & 58.2 & 25.6 \\
  GPT-4 (CoT) & 68.4 & \textbf{36.7} \\
  GPT-4 (ICL) & \textbf{69.9} & 34.9\\
  \hline
  Llama2-7B-chat (Default) & 11.9 & 10.2 \\
  Llama2-7B-chat (CoT) & \textbf{29.8} & \textbf{16.3} \\
  Llama2-7B-chat (ICL) & 10.4 & 15.3 \\
  \hline
  Gemma-7B-it (Default) & \textbf{45.7} & \textbf{31.6} \\
  Gemma-7B-it (CoT) & 36.1 & 17.3 \\
  Gemma-7B-it (ICL) & 33.1 & 31.6 \\
  \hline
  Qwen-7B-chat (Default) & \textbf{42.0} & 17.4 \\
  Qwen-7B-chat (CoT) & 40.2 & 22.4 \\
  Qwen-7B-chat (ICL) & 35.7 & \textbf{24.5} \\
  \hline
 \end{tabular}
 \caption{Accuracy of the music theory exercises of each model. All three modes of results are provided. \textit{Know.} means the music knowledge part and \textit{Reas.} means the music reasoning part. They are two subsets of which the former tests the models' memory of basic music concepts and the latter needs further reasoning and calculation to be completed. GPT-4's results come from ~\cite{yuan_chatmusician_2024}.}
 \label{tab:music_bench_results}
\end{table}

\begin{table}[ht]
    \centering
    \begin{tabular}{|c|c|c|c|c|c|c|c|}
    \hline
        \multirow{2}{*}{\textit{Type}} & \multirow{2}{*}{\textit{Model}} & \multicolumn{2}{|c|}{\textit{Inst. Fl.}} & \multicolumn{2}{c|}{\textit{Correct.}} & \multicolumn{2}{c|}{\textit{Reason.}}\\ 
    \cline{3-8}
        ~ & ~ & $\mu$ & $\sigma$ & $\mu$ & $\sigma$ & $\mu$ & $\sigma$ \\ 
        \hline
        \multirow{4}{*}{ME} & GPT-4 & 10.0 & 0.0 & 6.5 & 2.6 & 7.8 & 1.3 \\ \cline{3-8}
        ~ & Gemma & 8.2 & 2.1 & 5.1 & 2.8 & 7.4 & 3.2 \\ \cline{3-8}
        ~ & Llama2 & 7.8 & 1.9 & 4.7 & 2.8 & 4.7 & 2.4 \\ \cline{3-8}
        
        ~ & Qwen & 7.6 & 0.7 & 3.8 & 1.5 & 2.1 & 1.3 \\ \hline
        \multirow{4}{*}{MFE} & GPT-4 & 10.0 & 0.0 & 5.0 & 2.5 & 5.6 & 2.0 \\ \cline{3-8}
        ~ & Gemma & 3.5 & 3.6 & 2.1 & 2.2 & 2.9 & 2.1 \\ \cline{3-8}
        ~ & Llama2 & 5.4 & 1.8 & 3.2 & 2.8 & 4.3 & 2.3 \\ \cline{3-8}
        
        ~ & Qwen & 2.6 & 1.5 & 2.3 & 1.9 & 3.3 & 2.0 \\ \hline
    \end{tabular}
    \label{tab:comprehenssion_cot}
    \caption{The human assessment results of different LLMs on the understanding task. \textit{Inst. Fl.}, \textit{Correct.} and \textit{Reason.} respectively indicate to what extent the model follows the instructions, correctly answers the questions, and reasons like humans. $\mu$ and $\sigma$ respectively denote the average scores and the standard variance.}
\end{table}

\begin{table}[!ht]
    \centering
    \begin{tabular}{|l|l|l|l|l|l|}
    \hline
        \multirow{2}{*}{\textit{Type}} & \multirow{2}{*}{\textit{Model}} & \multicolumn{2}{|c|}{\textit{Inst. Fl.}} & \multicolumn{2}{|c|}{\textit{Condi.}} \\ 
    \cline{3-6}
        ~ & ~ & $\mu$ & $\sigma$ & $\mu$ & $\sigma$ \\ \hline
        \multirow{4}{*}{MFMC} & GPT-4  & 5.7 & 1.4 & 6.3 & 1.5 \\ \cline{3-6}
        ~ & Gemma & 4.0 & 1.8 & 4.6 & 2.2 \\ \cline{3-6}
        ~ & Llama2 & 4.3 & 1.6 & 4.3 & 2.3 \\ \cline{3-6}
        ~ & Qwen & 4.9 & 2.1 & 2.9 & 2.2 \\ \hline
        
        \multirow{2}{*}{MH} & GPT-4  & 6.5 & 3.5 & 5.5 & 2.5 \\ \cline{3-6}
        ~ & Gemma & 3.0 & 1.0 & 4.5 & 2.5 \\ \hline
        \multirow{2}{*}{CCG} & GPT-4 & 5.2 & 3.3 & 5.8 & 3.8 \\ \cline{3-6}
        ~ & Gemma & 1.6 & 1.0 & 1.3 & 0.8 \\ \hline
    \end{tabular}
    \label{tab:generation_wo_cot}
    \caption{The human assessment results of different LLMs on the generation task. \textit{Condi.} indicates to what extent the model contains the condition given in the instructions and ABC format.}
\end{table}

\subsection{Qualitative Results}


For human assessment, Table 3 
shows LLMs on ME and MFE tasks under the \textit{CoT} mode. We randomly sampled 40 examples of each task. In the instruction following question, GPT-4 demonstrates very good results, while other LLMs more or less can accomplish the tasks, indicating a certain level of capability. However, when it comes to the correctness, even GPT-4 finds it challenging to provide satisfactory answers to the prompts. When testing the logical reasoning of LLMs, the average scores indicate that all LLMs encounter difficulties in applying logical reasoning when answering questions step by step, leading to fundamental errors in music theory or illogical conclusions. This highlights the LLMs' limitation of involving music background knowledge. 


Table 4 
presents the results of human assessment we conducted on generative tasks. In addition to the results shown in the table, we also conducted an AB test based on Musicality. We find that the GPT-4 and Gemma-7B-it achieve comparable results in MFMCG task, while in other tasks GPT-4 always wins. This means Gemma-7B-it has a potential in creating high-quality symbolic music with limited model size.

As depicted in Figure~\ref{fig:responses_ABC_statistics}, on MH and CCG tasks, Qwen-7B-chat and Llama2-7B-chat struggled to effectively output correct ABC sequences to be rendered into audios. Therefore, for MH and CCG tasks, we only include the AB test results for GPT-4 and Gemma-7B-it. Despite GPT-4 achieving relatively better scores in generative tasks, it still falls far away from humans' expectations. Interestingly, beyond the data, LLMs' generative results occasionally exhibit instances of copying motifs provided in the prompt, as well as displaying unstructured harmonic repetitions or completely off-key notes. We believe that although LLMs can adhere to the ABC format condition provided in the prompt, their lack of musical information and knowledge makes it challenging to understand the high-level information within the condition, resulting in less satisfactory generated outcomes.


In terms of the results from subjective experiments, we identified a common issue prevalent in LLMs. Firstly, LLMs, apart from GPT-4, struggle to generate data in the correct ABC format with high probability, despite being able to provide a perfect answer when asked what ABC notation is. This phenomenon led us to speculate that while LLMs are trained extensively and comprehensively, LLMs can hardly understand all the information they have been exposed to and utilize them in different scenarios. Besides, LLMs can generate music in a seemingly appropriate ABC format in generative tasks, but what appears to be a correctly-formatted response is merely copying the prompt without grasping the semantic and structural information in the given condition.

\subsubsection{Multi-step Reasoning Analysis}

To better illustrate each model's reasoning capability when it is used to complete the music theory exercises, we provide an example of a question in the music theory exercises subset and step-by-step responses of the four models\footnote{https://github.com/SylviaZiyaZhou/LLMs\_music\_reasoning/blob/\\main/CoT\_music\_theory\_exercise\_all\_LLMs.pdf}. The question is about recognizing the interval property of an ABC sequence referring to a compound in a music sheet. From the responses, we can see that GPT-4 is the only model which can actually perform the calculation but still unable to understand the musical notes in the ABC notation. In the GPT-4's responses in the \textit{CoT} mode, "4", which is mistaken as "a fourth apart", should be a note duration. Accordingly, this mistake influences the whole reasoning process of the calculation of intervals. The response of Llama2-7B-chat also shows its incapability of involving correct music knowledge understanding of notes intervals in the reasoning process. What's more, Qwen-7B-chat even accidentally contains Chinese in the English text and Gemma-7B-it failed to recognize musical notes in the ABC sequence (see in the supplementary materials), although they can return the correct answer if they are merely asked about "the definition of note intervals". 

\begin{figure}
\centering
\includegraphics[width=1\columnwidth]{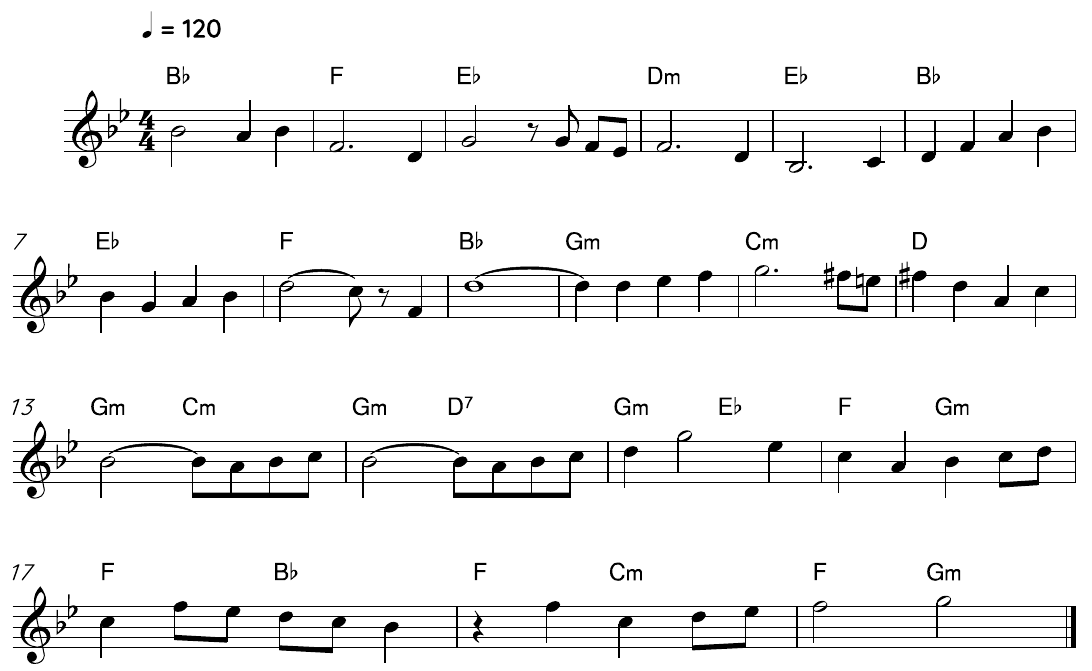}
\caption{Human composer's work for the chord-conditioned generation task.}
\label{fig:Captivating_Melody_human}
\end{figure}

Besides, the responses of generation tasks such as MFMC generation also have similar problems. In the CoT mode, we find all LLMs except GPT-4, are hard to follow the multi-step instructions and output music in a correct ABC format, so we only provide a GPT-4 response respectively in the raw text\footnote{https://github.com/SylviaZiyaZhou/LLMs\_music\_reasoning/blob\\/main/CoT\_music\_generation\_GPT4\_response.pdf} and music sheet\footnote{https://github.com/SylviaZiyaZhou/LLMs\_music\_reasoning/blob\\/main/Music\_Sheet\_of\_music\_generation\_GPT4\_response.pdf} form given the prompt in Figure~\ref{fig:Prompt_examples_generation}. Although GPT-4 can well understand the instructions in every step, it generates repetitive and simple rhythm without enough progression and variation, compared to the human composer's work in Figure~\ref{fig:Captivating_Melody_human}. 

\section{Conclusion and Discussion}
In conclusion, our experimental analysis highlights current LLMs' limitations in the realm of music understanding and generation, particularly from the perspective of song-level multi-step reasoning. These findings are crucial as they underline the challenges LLMs face when tasked with generating coherent and contextually rich musical compositions, which often require both complex sequential processing and creative fineness. From the human assessment results and the error analysis, we find that all these models failed to inject correct music theory and knowledge in the music understanding, reasoning and generation process. This knowledge generalization gap is analogous to the reversal curse problem illustrated in ~\cite{berglund2024reversal} where LLMs trained on “A is B” fail to learn “B is A”. Without making sure the fundamental concepts are correctly mentioned in the generated responses, it is hard to alleviate the LLMs' hallucination and guarantee the responses' quality. Therefore, it is significant to implement the knowledge augmentation module in the Supervised Fine-Tuning (SFT) stage to ensure the LLMs can reason based on correct music knowledge by curating more SFT data with enough knowledge-based contexts and practical reasoning processes. 

Specifically, several insights for the multi-step SFT dataset construction can be concluded from the process where professional musicians are asked to create music following the instructions. Firstly, more expert knowledge should be involved in the dataset construction to guarantee its quality. For example, in the chord-conditioned generation task in Chatmusician's dataset, the bass note sequence of the given chords does not conform to the musicians' expectation of the progression generally. 
Secondly, some conditions in the old one-step form are too lengthy and informative with limitations that the human composers feel difficult to follow. For example, when they are given an "AB" structure with two different motives in the MFMC task, all of them find hard to integrate two segments with different motives into a complete piece of music in an "AB" form. Therefore, it might not be reasonable to ask the LLMs to output a completely and well composed music in a one-step approach.

What's more, although four models are all claimed to be able to handle the input size from \textit{4K} to \textit{8K} tokens, which is much longer than the instructions in the dataset we used, they do not show their long-context processing advantages in the symbolic music domain. Our experimental results show that the widely-used CoT and ICL approaches are not always effective in improving the model's performance. In this way, more step-by-step learning strategies should be specifically developed for instruction-based symbolic music tasks by focusing on correctly answering music theory exercises, explicitly extracting motifs and implicitly extracting musical forms, and consistently following the conditions in the instructions.

\section{Acknowledgement}
The research was supported by Early Career Scheme (ECS-HKUST22201322), Theme-based Research Scheme (T45-205/21-N) from Hong Kong RGC, NSFC (No. 62206234), and Generative AI Research and Development Centre from InnoHK. Yinghao Ma is a research student at the UKRI Centre for Doctoral Training in Artificial Intelligence and Music, supported by UK Research and Innovation [grant number EP/S022694/1].

We would like to express our sincere gratitude to Jia Ding and Xiaoduan Li, professional musicians who majored in music composition at Central Conservatory of Music, for their valuable contributions and suggestions throughout the multi-step prompt engineering in conditioned generation tasks. 


\bibliography{ISMIR2024_template}

\begin{thebibliography}{10}
\providecommand{\url}[1]{#1}
\csname url@samestyle\endcsname
\providecommand{\newblock}{\relax}
\providecommand{\bibinfo}[2]{#2}
\providecommand{\BIBentrySTDinterwordspacing}{\spaceskip=0pt\relax}
\providecommand{\BIBentryALTinterwordstretchfactor}{4}
\providecommand{\BIBentryALTinterwordspacing}{\spaceskip=\fontdimen2\font plus
\BIBentryALTinterwordstretchfactor\fontdimen3\font minus \fontdimen4\font\relax}
\providecommand{\BIBforeignlanguage}[2]{{%
\expandafter\ifx\csname l@#1\endcsname\relax
\typeout{** WARNING: IEEEtran.bst: No hyphenation pattern has been}%
\typeout{** loaded for the language `#1'. Using the pattern for}%
\typeout{** the default language instead.}%
\else
\language=\csname l@#1\endcsname
\fi
#2}}
\providecommand{\BIBdecl}{\relax}
\BIBdecl

\bibitem{yuan_chatmusician_2024}
\BIBentryALTinterwordspacing
R.~Yuan, H.~Lin, Y.~Wang, Z.~Tian, S.~Wu, T.~Shen, G.~Zhang, Y.~Wu, C.~Liu, Z.~Zhou, Z.~Ma, L.~Xue, Z.~Wang, Q.~Liu, T.~Zheng, Y.~Li, Y.~Ma, Y.~Liang, X.~Chi, R.~Liu, Z.~Wang, P.~Li, J.~Wu, C.~Lin, Q.~Liu, T.~Jiang, W.~Huang, W.~Chen, E.~Benetos, J.~Fu, G.~Xia, R.~Dannenberg, W.~Xue, S.~Kang, and Y.~Guo, ``{ChatMusician}: {Understanding} and {Generating} {Music} {Intrinsically} with {LLM},'' Feb. 2024, arXiv:2402.16153 [cs, eess]. [Online]. Available: \url{http://arxiv.org/abs/2402.16153}
\BIBentrySTDinterwordspacing

\bibitem{garland1995math}
T.~H. Garland and C.~V. Kahn, \emph{Math and Music: Harmonious Connections.}\hskip 1em plus 0.5em minus 0.4em\relax ERIC, 1995.

\bibitem{wright2009mathematics}
D.~Wright, \emph{Mathematics and music}.\hskip 1em plus 0.5em minus 0.4em\relax American Mathematical Soc., 2009, vol.~28.

\bibitem{ding2024songcomposer}
S.~Ding, Z.~Liu, X.~Dong, P.~Zhang, R.~Qian, C.~He, D.~Lin, and J.~Wang, ``Songcomposer: A large language model for lyric and melody composition in song generation,'' \emph{arXiv preprint arXiv:2402.17645}, 2024.

\bibitem{liang_bytecomposer_2024}
\BIBentryALTinterwordspacing
X.~Liang, X.~Du, J.~Lin, P.~Zou, Y.~Wan, and B.~Zhu, ``{ByteComposer}: a {Human}-like {Melody} {Composition} {Method} based on {Language} {Model} {Agent},'' Mar. 2024, arXiv:2402.17785 [cs, eess]. [Online]. Available: \url{http://arxiv.org/abs/2402.17785}
\BIBentrySTDinterwordspacing

\bibitem{achiam2023gpt}
J.~Achiam, S.~Adler, S.~Agarwal, L.~Ahmad, I.~Akkaya, F.~L. Aleman, D.~Almeida, J.~Altenschmidt, S.~Altman, S.~Anadkat \emph{et~al.}, ``Gpt-4 technical report,'' \emph{arXiv preprint arXiv:2303.08774}, 2023.

\bibitem{team2024gemma}
G.~Team, T.~Mesnard, C.~Hardin, R.~Dadashi, S.~Bhupatiraju, S.~Pathak, L.~Sifre, M.~Rivi{\`e}re, M.~S. Kale, J.~Love \emph{et~al.}, ``Gemma: Open models based on gemini research and technology,'' \emph{arXiv preprint arXiv:2403.08295}, 2024.

\bibitem{touvron2023llama}
H.~Touvron, L.~Martin, K.~Stone, P.~Albert, A.~Almahairi, Y.~Babaei, N.~Bashlykov, S.~Batra, P.~Bhargava, S.~Bhosale \emph{et~al.}, ``Llama 2: Open foundation and fine-tuned chat models,'' \emph{arXiv preprint arXiv:2307.09288}, 2023.

\bibitem{bai2023qwen}
J.~Bai, S.~Bai, Y.~Chu, Z.~Cui, K.~Dang, X.~Deng, Y.~Fan, W.~Ge, Y.~Han, F.~Huang \emph{et~al.}, ``Qwen technical report,'' \emph{arXiv preprint arXiv:2309.16609}, 2023.

\bibitem{yu2023musicagent}
D.~Yu, K.~Song, P.~Lu, T.~He, X.~Tan, W.~Ye, S.~Zhang, and J.~Bian, ``Musicagent: An ai agent for music understanding and generation with large language models,'' \emph{arXiv preprint arXiv:2310.11954}, 2023.

\bibitem{yu2023reasoningsurvey}
F.~Yu, H.~Zhang, and B.~Wang, ``Nature language reasoning, a survey,'' \emph{arXiv preprint arXiv:2303.14725}, 2023.

\bibitem{hao2023reasoningLM}
S.~Hao, Y.~Gu, H.~Ma, J.~J. Hong, Z.~Wang, D.~Z. Wang, and Z.~Hu, ``Reasoning with language model is planning with world model,'' \emph{arXiv preprint arXiv:2305.14992}, 2023.

\bibitem{wei2022chain}
J.~Wei, X.~Wang, D.~Schuurmans, M.~Bosma, F.~Xia, E.~Chi, Q.~V. Le, D.~Zhou \emph{et~al.}, ``Chain-of-thought prompting elicits reasoning in large language models,'' \emph{Advances in neural information processing systems}, vol.~35, pp. 24\,824--24\,837, 2022.

\bibitem{kojima2022zeroshotlearners}
T.~Kojima, S.~S. Gu, M.~Reid, Y.~Matsuo, and Y.~Iwasawa, ``Large language models are zero-shot reasoners,'' \emph{Advances in neural information processing systems}, vol.~35, pp. 22\,199--22\,213, 2022.

\bibitem{chen2022program}
W.~Chen, X.~Ma, X.~Wang, and W.~W. Cohen, ``Program of thoughts prompting: Disentangling computation from reasoning for numerical reasoning tasks,'' \emph{arXiv preprint arXiv:2211.12588}, 2022.

\bibitem{wang2023plan}
L.~Wang, W.~Xu, Y.~Lan, Z.~Hu, Y.~Lan, R.~K.-W. Lee, and E.-P. Lim, ``Plan-and-solve prompting: Improving zero-shot chain-of-thought reasoning by large language models,'' \emph{arXiv preprint arXiv:2305.04091}, 2023.

\bibitem{zhang2023textgensurvey}
H.~Zhang, H.~Song, S.~Li, M.~Zhou, and D.~Song, ``A survey of controllable text generation using transformer-based pre-trained language models,'' \emph{ACM Computing Surveys}, vol.~56, no.~3, pp. 1--37, 2023.

\bibitem{chen2019sentiment}
H.~Chen, X.~Yi, M.~Sun, W.~Li, C.~Yang, and Z.~Guo, ``Sentiment-controllable chinese poetry generation.'' in \emph{IJCAI}, 2019, pp. 4925--4931.

\bibitem{dathathri2019plug}
S.~Dathathri, A.~Madotto, J.~Lan, J.~Hung, E.~Frank, P.~Molino, J.~Yosinski, and R.~Liu, ``Plug and play language models: A simple approach to controlled text generation,'' \emph{arXiv preprint arXiv:1912.02164}, 2019.

\bibitem{khalifa2020distributional}
M.~Khalifa, H.~Elsahar, and M.~Dymetman, ``A distributional approach to controlled text generation,'' \emph{arXiv preprint arXiv:2012.11635}, 2020.

\bibitem{tang2019topic}
H.~Tang, M.~Li, and B.~Jin, ``A topic augmented text generation model: Joint learning of semantics and structural features,'' in \emph{Proceedings of the 2019 conference on empirical methods in natural language processing and the 9th international joint conference on natural language processing (EMNLP-IJCNLP)}, 2019, pp. 5090--5099.

\bibitem{puduppully2019contentplan}
R.~Puduppully, L.~Dong, and M.~Lapata, ``Data-to-text generation with content selection and planning,'' in \emph{Proceedings of the AAAI conference on artificial intelligence}, vol.~33, no.~01, 2019, pp. 6908--6915.

\bibitem{ribeiro2020investigating}
L.~F. Ribeiro, M.~Schmitt, H.~Sch{\"u}tze, and I.~Gurevych, ``Investigating pretrained language models for graph-to-text generation,'' \emph{arXiv preprint arXiv:2007.08426}, 2020.

\bibitem{simon2022tattletale}
N.~Simon and C.~Muise, ``Tattletale: storytelling with planning and large language models,'' in \emph{ICAPS Workshop on Scheduling and Planning Applications}, 2022.

\bibitem{xie2024creating}
K.~Xie and M.~Riedl, ``Creating suspenseful stories: Iterative planning with large language models,'' \emph{arXiv preprint arXiv:2402.17119}, 2024.

\bibitem{zhang2024llm}
Z.~Zhang, M.~Rayhan, T.~Herda, M.~Goisauf, and P.~Abrahamsson, ``Llm-based agents for automating the enhancement of user story quality: An early report,'' \emph{arXiv preprint arXiv:2403.09442}, 2024.

\bibitem{berglund2024reversal}
L.~Berglund, M.~Tong, M.~Kaufmann, M.~Balesni, A.~C. Stickland, T.~Korbak, and O.~Evans, ``The reversal curse: Llms trained on "a is b" fail to learn "b is a",'' 2024.

\end{thebibliography}

\end{document}